\documentclass[prl,twocolumn,superscriptaddress,floatfix,letterpaper]{revtex4}
\usepackage{graphicx,amssymb,bm,amsfonts,amsmath}
\newcommand {\bisco}{Bi$_2$Sr$_2$CaCu$_2$O$_{8+\delta}$}

\newcommand {\SLbisco}{Bi$_2$Sr$_{1.6}$La$_{0.4}$Cu$_2$O$_{6+\delta}$}

\usepackage[bookmarks=false,pdfstartview=FitH]{hyperref}
\begin{document}

\title {A universal high energy anomaly in angle resolved photoemission spectra of high temperature superconductors -- possible evidence of spinon and holon branches}

\author{J. Graf} 
\affiliation{Materials Sciences Division, Lawrence Berkeley National Laboratory, Berkeley, CA 94720, USA}
\affiliation{Swiss Federal Institute of Technology Lausanne, CH-1015, Lausanne, Switzerland}
\author {G.-H. Gweon}
\affiliation {Department of Physics, University of California Berkeley, CA 94720, USA}
\affiliation {Department of Physics, University of California, Santa Cruz, California 95064, USA}
\author {K. McElroy}
\affiliation {Materials Sciences Division, Lawrence Berkeley National Laboratory, Berkeley, CA 94720, USA}
\author{S. Y. Zhou} \affiliation{Department of Physics, University of California Berkeley, CA 94720, USA} 
\author{C. Jozwiak} \affiliation{Department of Physics, University of California Berkeley, CA 94720, USA} 
\author {E. Rotenberg}
\affiliation {Advanced Light Source, Lawrence Berkeley National Laboratory, Berkeley, CA 94720, USA}
\author{A. Bill} \affiliation{Department of Physics, University of California Berkeley, CA 94720, USA}
\author {T. Sasagawa}
\affiliation {Department of Advanced Materials Science, University of Tokyo, Kashiwa, Chiba 277-8561, Japan}
\affiliation {CREST, Japan Science and Technology Agency, Saitama 332-0012, Japan}
\author {H. Eisaki}
\affiliation {AIST, 1-1-1 Central 2, Umezono, Tsukuba, Ibaraki, 305-8568, Japan}
\author {S. Uchida}
\affiliation {Department of Physics, University of Tokyo, Yayoi, 2-11-16 Bunkyoku, Tokyo 113-8656, Japan}
\author {H. Takagi}
\affiliation {Department of Advanced Materials Science, University of Tokyo, Kashiwa, Chiba 277-8561, Japan}
\affiliation {CREST, Japan Science and Technology Agency, Saitama 332-0012, Japan}
\affiliation {RIKEN (The Institute of Physical and Chemical Research), Wako 351-0198, Japan}
\author{D.-H. Lee}
\affiliation{Materials Sciences Division, Lawrence Berkeley National Laboratory, Berkeley, CA 94720, USA} \affiliation{Department of Physics, University of California Berkeley, CA 94720, USA}
\author{A. Lanzara} \email{alanzara@lbl.gov}
\affiliation{Materials Sciences Division, Lawrence Berkeley National Laboratory, Berkeley, CA 94720, USA} \affiliation{Department of Physics, University of California Berkeley, CA 94720, USA}

\date {\today}

\begin{abstract}
A universal high energy anomaly in the single particle spectral function is reported in three different families of high temperature superconductors by using angle-resolved photoemission spectroscopy. As we follow the dispersing peak of the spectral function from the Fermi energy to the valence band complex, we find dispersion anomalies marked by two distinctive high energy scales, $E_1\approx 0.38$ eV and $E_2\approx 0.8$ eV\@.
$E_1$ marks the energy above which the dispersion splits into two branches.  One is a continuation of the near parabolic dispersion, albeit with reduced spectral weight, and reaches the bottom of the band at the $\Gamma$ point at $\approx 0.5$ eV. The other is given by a peak in the momentum space, nearly independent of energy between $E_1$ and $E_2$.  Above $E_2$, a band-like dispersion re-emerges.  We conjecture that these two energies mark the disintegration of the low energy quasiparticles into a spinon and holon branch in the high T$_c$ cuprates. 
\end{abstract}
\pacs{74.72.-h, 74.25.Jb, 79.60.-i}

\keywords{cuprates; strongly correlated electron systems; photoemission; ARPES; high-temperature superconductors; kink}
\maketitle

\paragraph*{}
\label{sec:INTRODUCTION}

Understanding how doped oxygen holes are transported in the environment of antiferromagnetically coupled copper spin is one of the most fundamental problems in the field of high temperature superconductivity. In 1988 Zhang and Rice \cite{Zhang88} proposed that the doped holes in the oxygen 2p$\sigma$ orbitals form singlets with the spins of the neighboring coppers. The resulting charge-e and spin-0 object is called the Zhang-Rice singlet (ZRS).  As the ZRS moves through the $CuO_2$ plane, the copper spins get rearranged. As a result, the ZRS couples very strongly to the antiferromagnetic environment. Remarkably as a consequence of such strong coupling, quasiparticles emerge at low energies. This is evidenced by the sharp nodal quasiparticle peaks seen in angle-resolved photoemission (ARPES) of almost all cuprate compounds \cite{Damascelli03,Campuzano04}. In simple physical terms a quasiparticle is a composite object made of a ZRS and a S=1/2 copper spins.  It is widely believed that, at sufficiently low temperatures, superconducting pairing occurs between these quasiparticles giving rise to the high temperature superconducting state.  Thus a microscopic understanding of the pairing mechanism of high Tc superconductors requires an in-depth understanding of how a ZRS is dressed into a quasiparticle. 

Here we present the first systematic study of the evolution of the ARPES spectral function from the Fermi level ($E_F\equiv0$) to the valence band complex (at energy $\approx 1$ eV \cite{Wells89}) for three different families of high temperature superconductors.  Our results provide a surprising new experimental understanding on the important quasiparticle formation process discussed above.  Specifically, by covering a much broader energy range than typically studied \cite{Damascelli03}, we have identified anomalies in the ARPES spectra occurring at two universal high energy scales, $E_1\approx 0.38$ eV and $E_2\approx 0.8$ eV from $E_F$.  We conjecture that these two energies mark the threshold for the disintegration of the low-energy quasiparticles at two different binding levels.   

\paragraph*{}
\label{sec:EXPERIMENTAL TECHNIQUE}
ARPES data have been collected at the Advanced Light Source, beamlines 7.0.1, 10.0.1 and 12.0.1. for three different families of hole-doped cuprates:
single layer \SLbisco\ (Bi2201), double layer \bisco\ (Bi2212)
and Pb-doped Bi2212 (Pb2212) and for
several doping values.  The data presented here were measured at least in both
the first and the second BZs and along the two polarization p$_a$ and p$_b$ as shown in Figure 1. The double layer Bi2212 data were collected at 52eV for the UD ($T_c$=64K) and OPT doped ($T_c$=91K) sample, and 33eV and 65eV for OD  ($T_c$=80K). The OD Pb2212 ($T_c$=65K) was measured at 55, 60 and 75 eV and the OPT doped Bi2201 ($T_c$=32K) at 33 eV.  Unless specified otherwise, all the data reported were measured at 25K.

\paragraph*{}
\label{sec:RESULTS}
Figure 1 shows the ARPES intensity map as a function of energy and momentum in the (0, 0)-$(\pi,\pi)$ direction for an (a) underdoped (UD), (b) optimally-doped (OPT), and (c) over-doped (OD) Bi2212.
In all panels two main features are apparent: a high intensity feature (yellow) at low energy (widely
studied in the literature \cite{Damascelli03,Campuzano04}), and a weaker intensity feature (red) at high energy. The high energy feature, ``waterfall''-like feature, is the main focus of this paper. Given the large energy span of Figure 1, the ``kink'' at $E_0\approx 0.06$ eV (gray arrows pointing to the right) is a very subtle feature. Aside from the kink, the low-energy dispersion can be well fitted by a single tight
binding band (dotted gray line in panel a) \cite{Damascelli03,Gweon04}.
Surprisingly, as the parabolic band reaches $E_1 = 0.38 \pm 0.07$ eV, at momentum around $(\pi/4, \pi/4)\frac{1}{a}$ (gray arrows pointing to the left), the dispersion suddenly undergoes a steep down-turn accompanied by a substantial drop of the ARPES intensity. As shown in Figure 1, the overall feature of this anomaly is nearly independent of doping.

\begin{figure}[!t] \includegraphics[width=8cm]{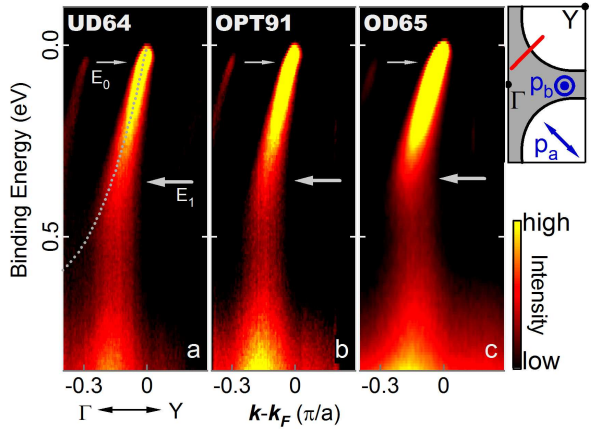}
\caption{\label{fig:1}(Color online). ARPES intensity maps of Bi2212 samples for three different
doping values. Data for the overdoped sample are in the normal state, 100K.
The location of the cut is shown by a red line in the BZ diagram on the right side, along with the different ARPES geometries used here. The gray dotted line is the dispersion obtained from a tight binding fit \cite {Gweon04} up to energy 0.3 eV\@.
}\end{figure}

In Figure 2 we present selected raw EDCs (energy distribution curves, energy cuts at constant momentum; panel a) and MDCs (momentum distribution curves, momentum cuts at constant energy; panel b) for the overdoped Bi2212. We show results for the overdoped sample, making a strong case that pseudogap \cite{Loeser96,Ding96}, disorder and inhomogeneity \cite{Pan01,Thurston89,Lee04} can be ruled out as possible origin. However, similar behaviors are observed for all doping values we studied. 

\begin{figure}[!b]\includegraphics[width=7.5cm]{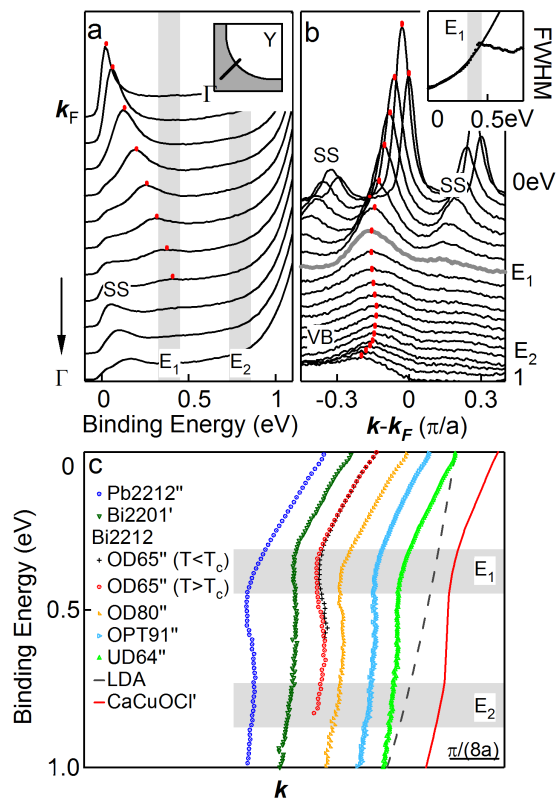}
\caption{\label{fig:2}(Color online). (a,b) EDCs from $\bm{k_F}$ (top curve) toward $\Gamma$ and MDCs for the OD Bi2212. The spectra are vertically shifted for an easy view. The small peaks on the right and left of the main MDC peak (red symbols) are due to the superstructure (SS). At $\approx 1$eV, the valence band (VB) can also be distinguished. The inset shows the FWHM of the MDC peak as a function of energy. (c) MDC dispersion, horizontally shifted for an easy view, for several compounds and doping. The Ca$_2$CuO$_2$Cl$_2$ dispersion \cite{Ronning05} is shifted in energy by $0.45$ eV to account for the energy gap. The OD-Bi2212 data are shown both above, 100K, and below Tc, 25K.  The prime and double prime stand for data taken in the first and second BZ respectively. The dashed line is the LDA band dispersion \cite{Lin06}}
\end{figure}

Panels a and b show that the behaviors of MDC and EDC peak become completely different as $E_1$ is reached, exposing the full view of the anomaly. The EDC peaks shown in panel a disperse in a simple manner and, as the momentum moves away from the Fermi momentum, the peak gets broader and weaker losing rapidly its strength as it approaches $E_1$ toward the high energy feature, well before it reaches the zone center, $\Gamma$. This is consistent with the sudden decrease of the ARPES intensity observed at $E_1$ in Figure 1.  In contrast, the raw MDCs in panel b show a well defined peak over the full energy range. For energy $\lessapprox E_1$, the MDC peak disperses in a consistent fashion with the EDC dispersion, and as it reaches $E_1$ it suddenly stops moving,
and becomes almost energy independent all the way up to $E_2$ and pinned at $\approx (\pi/4, \pi/4)\frac{1}{a}$. As the energy increases beyond $E_2$, the MDC peak starts dispersing again. It is surprising that a well defined MDC peak can still be identified within this large energy range. 
A further tracking of the MDC dispersion, well above $E_2$, is made difficult due to the strong valence band complex dominating the ARPES signal, as seen by the strong rise of the intensity on the right end side of panel a \cite{Wells89}.
Interestingly, the MDC peak width stops increasing at $E_1$ and shows a small {\em decrease} (inset of panel b).

From now on, we will refer collectively to the anomalies at $E_1$ and $E_2$ as ``high energy anomaly''. 
In figure 2c we report MDC dispersions for different materials,  various doping values, and different temperatures.  It is clear that the overall features of the high energy anomaly and their energy and momentum locations are a universal feature in all the materials studied, from heavily overdoped to undoped compound, and do not show any substantial change going from the superconducting to the normal state.  
Interestingly, beyond $E_2$ the MDC dispersion is in a reasonable agreement with the LDA prediction (panel c).  This gives an estimate of the energy scale of the measured band to be $\approx 1.3-1.4$ eV.

\begin{figure*}[!t] \includegraphics[width=15cm]{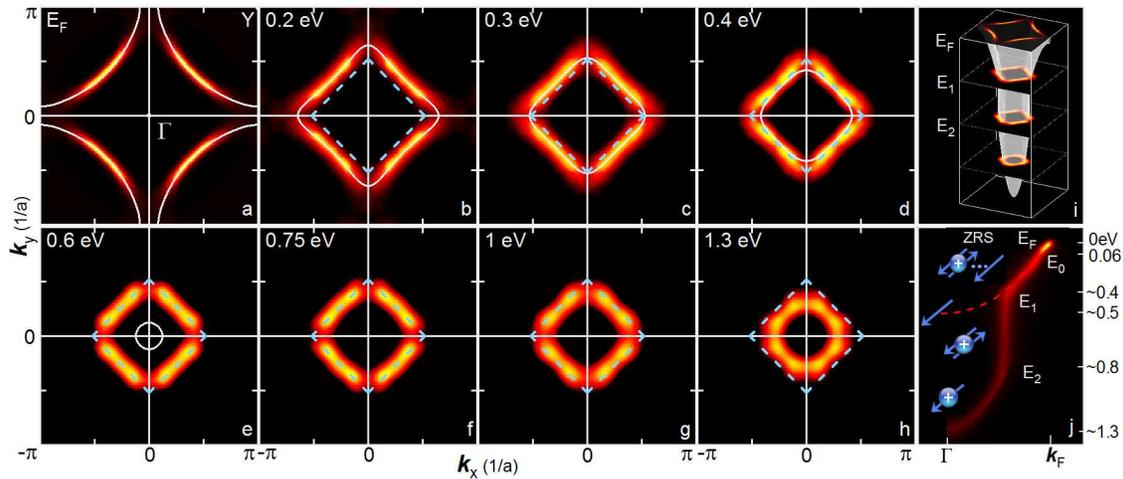}
\caption{\label{fig:3}(Color online). (a-h) Maps of the ARPES intensity in the momentum space at increasing energies for the Pb-Bi2212 sample. Data were taken in the second BZ and symmetrized according to the tetragonal symmetry. The color scale is normalized independently for each cut. The white solid lines correspond to the tight binding fit and the dashed light blue diamond indicates the characteristic geometry of the high energy anomaly. (i) Three dimensional plot of the ARPES intensity as a function of energy and in-plane momentum. (j) Our proposed scenario for the high energy anomaly. 
}\end{figure*}

In Figure 3 the full momentum space information about the high energy anomaly is summarized.
Panels a-h show the momentum space distribution of the ARPES intensity at several different energies from E$_F$ to $1.3$ eV. Representative data from overdoped Pb-Bi2212 are shown.
From $E_F$ to $\approx E_1$ (panels a-c) the ARPES contour and the tight binding fit (white solid line) are in very good agreement. This region corresponds to the low energy region, below $E_1$, of Figures 1 and 2. As the energy increases, the $E_1$ anomaly is marked unambiguously by the strong departure of the data from the tight binding fit and, over a wide energy region from $E_1$ to $E_2$ (panels d-f), the main ARPES intensity is ``pinned'' at the boundary of a diamond (light blue dashed line) whose four corners are located near $(\pm \pi/2a,0)$ and $(0,\pm\pi/2a)$.
We caution readers that despite the color scale the spectral weight within this diamond is {\it not} strictly zero (Figure 2).
When the energy increases beyond $E_2$ (panels g,h), the ARPES intensity starts moving again toward $\Gamma$, consistently with the resumed dispersion of the MDC peak discussed in Figure 2. 
This is a common feature of all the materials reported here and of optimally doped La$_{1.64}$Eu$_{0.2}$Sr$_{0.16}$CuO$_4$ (Eu-LSCO) \cite{Graf-unp}. 

We have searched for, but did not find, a similar anomaly in more conventional materials such as GaAs (a simple band insulator), K$_{0.9}$Mo$_6$O$_{17}$, SmTe$_3$ \cite{Gweon98} and CeTe$_2$ \cite{GarciaPriv} (quasi-two-dimensional metals with conventional charge density wave orders), and graphite (a quasi-two-dimensional semi-metal) \cite{Zhou05}. However, a similar high energy anomaly is also seen in several ruthenate compounds \cite{DenlingerPriv} suggesting that it might be an intrinsic feature of Mott physics.

\begin{figure}[!t] 
\includegraphics[width=8.8cm]{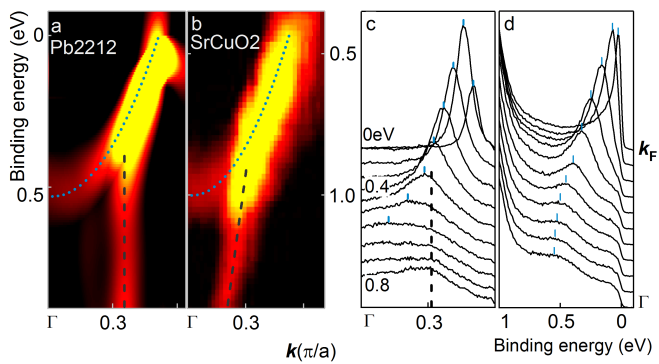}
\caption{\label{fig:4}(Color online). Second derivatives of ARPES intensity maps along the nodal direction of Pb2212 (a) and SrCuO$_2$ (b)\cite{Kim06}.  Only negative values of second derivatives are shown, to trace peaks but not dips.  Sum of the second momentum-derivative and the second energy-derivative is shown at each point.  (c,d) MDCs from E$_F$ (top curve) to 0.8eV and EDCs from $\bm{k_F}$ (top curve) to $\Gamma$ for the the data in panel (a). The MDCs and EDCs are vertically shifted for an easy view. The blue dotted line and the black dashed line highlight the proposed spinon and holon dispersions, respectively.}
\end{figure}

We note that this peculiar high energy behavior described in Figures 1-3 can not be explained by well known ARPES matrix element effect \cite{Bansil99}. The matrix element is strongly sensitive to the experimental settings in contrast to the high energy anomaly, e.g.~predicting \cite {BansilPriv} that the intensity near the $\Gamma$ point is {\em enhanced} in the second BZ, while it does predict an intensity depression in the first BZ\@.  Instead, we find a similar high energy behavior independently of the BZ, the photon energy (25, 33, 43, 52, 55, 59, 75, 90, 100, 130 and 150 eV), or the light polarization settings (along the Cu-O bonds, along the Cu-Cu bonds, and normal to the CuO$_2$ planes).

One possible explanation of the observed high energy anomaly is in terms of coupling to a bosonic mode along a similar line as the low energy kink.  However we believe this picture is unlikely since so far there is no known mode occurring at these energies and robust over the whole doping range, from zero doping to high over-doping.

A far more general possibility is that the data presented here show the disintegration of the low energy quasiparticle.
In this view, we propose that the dispersive band in the energy range from $E_0$ to $E_1$ is the signature of a composite object, as schematically represented in Figure 3j, made of a ZRS bound to a Cu spin 1/2.
This composite particle has quantum number S=1/2 and charge e and is consistent with that of a photohole. 
At lower energies this composite object is further dressed by phonons and low energy collective spin excitations to become the ultimate quasiparticle in the energy range between $E_F$ and $E_0$. 
The characteristics of the fermionic composite particle existing between $E_0$ and $E_1$ is the broad EDC and MDC peaks and their mutually consistent dispersions.
At $E_1$ this composite object breaks down into a ZRS and a copper spin.
The former is referred as a ``holon'' and the latter a ``spinon'' in the theory literature of high T$_c$.
Experimentally this is seen as the sudden loss of spectral intensity of the broad dispersive ARPES peaks at $E_1$. 
In the energy range between $E_1$ and $E_2$ the photoemission spectrum is the convolution of those of a spinon and a holon. In principle, the spectral features associated with both excitations can be observed by ARPES \cite{Kim06}.

Recently by tuning the energy (60 eV) and the polarization (out-of-plane) of the photon, we have observed for the first time a new faint dispersive feature (blue markers in Figure 4a) whose bottom occurs at $\approx$ 0.5 eV at the $\Gamma$ point. 
The corresponding peak of this new branch can be observed both in the MDCs and also in the EDCs spectra (panel c,d). 
We interpret this new feature as the spinon branch while the waterfall feature as the holon branch. The two peaks in the MDCs (panel c) in the energy range above 0.4 eV and below $\approx$ 0.5 eV represent the two branches.
This interpretation is supported by the striking similarity between our data (panel a) and the recent photoemission result of the spinon and holon branches of the one dimensional cuprate (panel b) \cite{Kim06}.  Note, however, that, while the comparison between panels a,b is quite appealing, not only qualitatively but also quantitatively, in our case two peaks can be observed only in MDCs but not in EDCs, due to the puzzling near-vertical nature of the holon dominated branch (waterfall). Finally, $E_2$ is the energy where the ZRS disintegrates into a bare oxygen hole and a copper spin. This reappearance of the fermionic oxygen hole explains the re-emergence of a band-like dispersion.
Obviously, more investigations would be necessary to test our conjecture further.

In conclusion, we have reported for the first time a universal high energy anomaly in the ARPES spectra of different families of high temperature superconductors, identified by a sudden change in the dispersion of the main spectral peak. This phenomenon is robust under the change of doping, as well as chemical composition. We conjecture that the high energy anomaly provides the long-sort-after evidence of spin charge separation in the high T$c$ compounds.

\begin{acknowledgments}
We would like to thank P. W. Anderson, A. Bansil, A. Bianconi, C. Di Castro, C. Castellani, S. Chakraverty, J.E. Hirsch, T. Egami, M. Jarrell, S. Kivelson, R.S. Markiewicz, A. Macridin, V. Oganesyan, P. Phillips, A. Perali and S. Sahrakorpi for useful discussions and A. Bostwick and A.V. Fedorov for experimental help. This work was supported by the Director, Office of Science, Office of Basic Energy Sciences, Division of Materials Sciences and Engineering, of the U.S. Department of Energy under Contract No. DE-AC03-76SF00098, and by the National Science Foundation through Grant No. DMR-0349361.
ALS is operated by the DOE's Office of BES, Division of Materials Science, under Contract No. DE-AC03-76SF00098.
\end{acknowledgments}


\end{document}